\documentclass[twocolumn,showpacs,preprintnumbers,pra,amsmath,amssymb]{revtex4}

\usepackage{graphicx}
\usepackage{bm}
\begin{document}
\title{Possibility of upper-bounding dimension of quantum states device-independently}

\author{Won-Young Hwang \footnote{Email: wyhwang@jnu.ac.kr}}

\affiliation{$^{1}$Department of Physics Education, Chonnam National University, Gwangju 500-757, Republic of Korea}

\begin{abstract}
We discuss possibility of upper-bounding dimension of quantum states device-independently. Provided that the states are pure, it is possible to generate certain four states whose dimension is bounded by two.
\end{abstract}
\pacs{03.67.-a, 03.67.Dd}


\maketitle
\section{Introduction}
Two-dimensional quantum state, or quantum bit (qubit), has been an important element in quantum information processing.   Theoretically, qubit is the most simple one to deal with.
In practice, however, it is difficult to generate perfect qubits because no-device is perfect. The difficulty is not a serious problem in most cases if degree of imperfection is small. However, in quantum key distribution (QKD) it is a serious problem because effect of the imperfection can be cleverly amplified by an eavesdropper \cite{Yue96,Sca09}. In fact, problems due to unavoidable imperfections have been major issues in practical QKD \cite{Sca09}.

Recently, semi-device-independent QKD has been proposed, whose proviso is that dimension of source is bounded by two, that is, source is qubit \cite{Paw11}. Moreover, it has been shown \cite{Yin13,Yin14} that currently the most practicable one, the measurement-device-independent QKD gets full security provided that quantum states are two-dimensional. Thus an interesting issue is how to upper-bound dimension of quantum states device-independently. Recently studied ``dimension witness \cite{Bru08}" is a one which lower-bounds dimensions.

On the other hand, it has been shown with use of no-signaling principle that even with un-characterized devices it is possible to generate four quantum states for which the unambiguous state discrimination \cite{Iva87} is not possible \cite{Hwa12}. The result implies that, if the states are pure, dimension of  the states is bounded by three, because unambiguous state discrimination is always possible for linearly independent states \cite{Che98}. In this paper, we describe the prescription to generate the four states. Then we show that dimension of the generated states is bounded by two if the states are pure.
\section{to generate the four states}
Let us recall a method to generate a set of qubits using Bell states. The set of qubits, $\{|0\rangle, |1\rangle, |+\rangle, |-\rangle\}$, is the most widely used one in QKD.
Here, $|0\rangle$ and $|1\rangle$ are orthonormal states and $|\pm\rangle = (1/\sqrt{2}) (|0\rangle \pm |1\rangle)$. Assume that a sender, normally called Alice, wants to generate the states. If Alice has ideal Bell state and can perform ideal measurements, she can generate the ideal states;
Alice has a Bell state
\begin{equation}
|\phi^+\rangle= \frac{1}{\sqrt{2}} (|0\rangle_{A_1} |0\rangle_{A_2} + |1\rangle_{A_1} |1\rangle_{A_2}),
\label{1}
\end{equation}
where $A_1$ and $A_2$ denote two separated boxes in which corresponding quantum states are kept. Assume that $M_0$ ($M_1$) is a measurement composed of $\{|0\rangle \langle 0|, |1\rangle \langle 1|\}$ ($\{|+\rangle \langle +|, |-\rangle \langle -|\}$).
Alice chooses a value $i$ ($i=0,1$) with equal probability. She performs $M_i$ on quantum state in box $A_1$. Then, in box $A_2$, each one of the four qubit states are generated with equal probability. However, in practice, every component has unavoidable imperfections. Hence the actually generated state deviates from the ideal states. As a result, the dimension of the actual state also deviates from two.

However, it is possible to generate states whose dimension is strictly bounded by two even with un-characterized imperfect devices, provided that the states are pure. In brief, Alice follows the same procedures with imperfect real devices as she did with the ideal ones: She prepares a bipartite quantum state which we denote by $\tilde{\rho}$.
Then one part (the other one) of the state is kept into a box with a label $A_1$ (a label $A_2$). She randomly chooses a binary bit $i$. Alice performs a measurement $\tilde{M_i}$ on quantum state in box-$A_1$ obtaining an outcome $j$ between $0$ and $1$. She attaches a label $(i,j)$ on box  $A_2$.
If $\tilde{\rho}= |\phi^+\rangle$ and $\tilde{M_i}= M_i$, the states in box-$A_2$ with a label $(0,0), (0,1), (1,0), (1,1)$, are $|0\rangle, |1\rangle, |+\rangle, |-\rangle$, respectively. However, for any state  $\tilde{\rho}$ and any measurement $\tilde{M_i}$ , our argument below for two-dimensionality is valid. However, in order that the generated states be close to the specified qubits, the state and measurement should be close to the Bell state and $M_i$.
\section{bounded by two dimension}
Let $p(j|i)$ denote a probability that an outcome $j$ is obtained when Alice performed a measurement $\tilde{M_i}$. Here  $\sum_j p(j|i)= 1$ for each $i$. Here the state in box-$A_2$ with a label $(i,j)$ is supposed to be pure and denoted by $|i,j\rangle$. Clearly, the density operator of the states in box-$A_2$, which is $tr_{A_1}(|\phi^+\rangle \langle\phi^+|)$, must not depend on $i$ value \cite{Nie00}.
\begin{eqnarray}
 && p(0|0) |0,0\rangle \langle 0,0|  + p(1|0) |0,1\rangle \langle 0,1| \nonumber\\
&=& p(0|1) |1,0\rangle \langle 1,0| + p(1|1) |1,1\rangle \langle 1,1|.
\label{2}
\end{eqnarray}
Note that Eq. (\ref{2}) is satisfied rigorously even if the states are prepared with un-characterized imperfect devices as long as Alice followed the procedures for preparation. Otherwise, superluminal communication become possible violating a fundamental principle in nature. The only case when Eq. (\ref{2}) can be violated is that Alice fails to follow the procedure, e.g., an additional information carrier was delivered to the box-$A_2$ after measuring state in box-$A_1$. We assume that there was no such hidden signaling.
Without the reasonable assumption, cryptographic task is not even possible. The assumption is made in all analysis of practical and device-independent QKD \cite{Sca09,Aci06}.

It is easy to see that the states are of two-dimensional. We show that two states of them are linear combination of the other two states. Let us introduce a state $|(1,1)^{\perp} \rangle= |1,0\rangle - \langle 1,1| 1,0 \rangle |1,1 \rangle$,
which is orthogonal to $|(1,1)\rangle$. By Applying the operator in Eq. (\ref{2}) on the state $|(1,1)^{\perp}\rangle$, we get
\begin{eqnarray}
 && p(0|0) \langle 0,0|(1,1)^{\perp}\rangle |0,0\rangle  + p(1|0) \langle 0,1|(1,1)^{\perp}\rangle |0,1\rangle \nonumber\\
&=& p(0|1) (1-|\langle 1,1| 1,0 \rangle |^2) |1,0\rangle.
\label{3}
\end{eqnarray}
Similarly, we obtain
\begin{eqnarray}
 && p(0|0) \langle 0,0|(1,0)^{\perp}\rangle |0,0\rangle  + p(1|0) \langle 0,1|(1,0)^{\perp}\rangle |0,1\rangle \nonumber\\
&=& p(1|1) (1-|\langle 1,1| 1,0 \rangle |^2) |1,1\rangle.
\label{4}
\end{eqnarray}
In the case when  $1-|\langle 1,1| 1,0 \rangle |^2\neq 0$, each of the two states $|1,0\rangle$ and $|1,1\rangle$ is linear combination of $|0,0\rangle$ and $|0,1\rangle$. This implies that the dimension is bounded by two.
Let us consider the other case when $1-|\langle 1,1| 1,0 \rangle |^2= 0$. We have $|1,1\rangle= e^{i\theta} |1,0 \rangle$ by Cauchy-Schwarz inequality, where $\theta$ is a real number. Thus the density operator corresponds to a pure state. This means that the state $\tilde{\rho}$ is a product state \cite{Nie00}. Then the four states become identical, dimension of which is clearly bounded by two.
\section{discussion and conclusion}

Because the state $\tilde{\rho}$ fluctuates each time in real implementation, so does the density operator for the state in box-$A_2$. So we cannot say that dimension of all states in box-$A_2$ generated during whole protocol is bounded by two. What is guaranteed is that dimension of four states that could have been potentially generated for each measurement is bounded, provided that the states are pure. However, what we need in QKD is the difficulty to discriminate between the four states that could have been generated potentially, which is provided by low dimensionality of the four states each time.

In conclusion, we discussed a possibility to generate a set of four state whose dimension is bounded by two, provided that the states are pure. This might be a clue to find how to upper-bound a state without any proviso.

\section*{Acknowledgement}
This study was supported by Basic Science Research Program through the National Research Foundation of Korea (NRF) funded by the Ministry of Education, Science and Technology (2010-0007208).


\begin{references}
\bibitem{Yue96} H. P. Yuen, Quantum Semiclassical Opt. {\bf 8}, 939 (1996).
\bibitem{Sca09} V. Scarani, H. Bechmann-Pasquinucci, N. J. Cerf, M. Dusek, N. L\"utkenhaus, M. Peev, Rev. Mod. Phys. {\bf81}, 1301 (2009).
\bibitem{Paw11} M. Pawlowski and N. Brunner, Phys. Rev. A {\bf 84}, 010302 (2011).
\bibitem{Yin13} Z.-Q. Yin et al., Phys. Rev. A {\bf 88}, 062322 (2013).
\bibitem{Yin14} Z.-Q. Yin et al., arxiv:1407.1924v2[quant-ph].
\bibitem{Bru08} N. Brunner et al., Phys. Rev. Lett. {\bf 100}, 210503 (2008).
\bibitem{Iva87} I.D. Ivanovic, Phys. Lett. A {\bf 123}, 257 (1987).
\bibitem{Hwa12} W.-Y. Hwang, J. Korean Phys. Soc. {\bf 60}, 1837 (2012).
\bibitem{Che98} A. Chefles, Phys. Lett. A {\bf 239}, 339 (1998).
\bibitem{Nie00} M. A. Nielsen and I. L. Chuang, {\it Quantum Computation and Quantum Information}, (Cambridge
            Univ. Press, Cambridge, U.K., 2000.)
\bibitem{Aci06} A. Ac\'in, N. Gisin, and L. Masanes, Phys. Rev. Lett. {\bf 97}, 010503 (2006).
\bibitem{Li11} H.-W. Li, Z.-Q. Yin, Y.-C. Wu, X.-B. Zou, S. Wang, W. Chen, G.-C. Guo, and Z.-F. Han, Phys. Rev. A {\bf 84}, 034301 (2011).
\end{references}
\end{document}